\documentstyle[fleqn,epsfig]{article} 

\newcommand{\AmS}{{\protect\the\textfont2
  A\kern-.1667em\lower.5ex\hbox{M}\kern-.125emS}}

\newcommand {\delm}{$\Delta m^{2}$\hspace{1mm}} 
\newcommand {\sqsin}{$\sin^{2} 2\theta$\hspace{1mm}}
\newcommand {\delmns}{$\Delta m^{2}$} 
\newcommand {\sqsinns}{$\sin^{2} 2\theta$}
\newcommand {\sqdm}{$\sin^{2} 2\theta,\Delta m^{2}$\hspace{1mm}}

\newcommand {\ltenloe}{$\log_{10}L/E$\hspace{1mm}}
\newcommand {\numutonutau}{$\nu_\mu \rightarrow \nu_\tau$\hspace{1mm}}

\newcommand{\enu}{\mbox{$E_\nu$}}
\newcommand{\numu}{\mbox{$\nu_{\mu}$}\hspace{1mm}}                   
\newcommand{\nue}{\mbox{$\nu_{e}$}\hspace{1mm}}                      
\newcommand{\nutau}{\mbox{$\nu_{\tau}$}\hspace{1mm}}                 

\newcommand{\anumu}{\mbox{$\overline{\nu}_{\mu}$}}      
\newcommand{\mup}{\mbox{$\mu^{+}$}}               
\newcommand{\mum}{\mbox{$\mu^{-}$}}               

\title{
\vskip 30pt
{\bf  Neutrino Oscillation Effects in Soudan-2 
      Upward-stopping Muons } 
}

\author{ 
W.W.M.~Allison$^3$, G.J.~Alner$^4$, 
D.S.~Ayres$^1$, G.D.~Barr$^3$, W.L.~Barrett$^6$,\\ 
P.M.~Border$^2$, J.H.~Cobb$^3$, D.J.A.~Cockerill$^4$, 
H.~Courant$^2$, D.M.~Demuth$^2$, \\
T.H.~Fields$^1$, H.R.~Gallagher$^5$, 
M.C.~Goodman$^1$, T.~Kafka$^5$, \\
S.M.S.~Kasahara$^2$, P.J.~Litchfield$^2$, 
W.A.~Mann$^5$, M.L.~Marshak$^2$, \\
W.H.~Miller$^2$,  L.~Mualem$^2$, J.K.~Nelson$^{2,a}$,
A.~Napier$^5$, W.P.~Oliver$^5$, \\
G.F.~Pearce$^4$, E.A.~Peterson$^2$, D.A.~Petyt$^2$, 
K.~Ruddick$^2$, M.~Sanchez$^{5,b}$, \\
J.~Schneps$^5$, A.~Sousa$^5$, J.L.~Thron$^{1,c}$, N.~West$^3$\\ 
\\ 
$^1${\it Argonne National Laboratory, Argonne, IL 60439}\\ 
$^2${\it University of Minnesota, Minneapolis, MN 55455}\\ 
$^3${\it Department of Physics, University of Oxford, Oxford OX1 3RH, UK}\\ 
$^4${\it Rutherford Appleton Laboratory, Chilton, Didcot, Oxfordshire 
 OX11 0QX, UK}\\ 
$^5${\it Tufts University, Medford, MA 02155}\\ 
$^6${\it Western Washington University, Bellingham, WA 98225}\\ 
\\
$^a${\small Now at the College of William and Mary, Williamsburg, VA 23187.}\\ 
$^b${\small Now at Harvard University, Cambridge, MA 02138.}\\ 
$^c${\small Now at Los Alamos National Laboratory, Los Alamos, NM 87545.}\\ 
} 

\begin{document} 

\maketitle 

\thispagestyle{empty} 

\begin{abstract} 
\normalsize 

Upward-going stopping muons initiated by atmospheric \numu~and \anumu~ 
interactions in the rock below the Soudan 2 detector have been isolated,
together with a companion sample of neutrino-induced single muons,
created within the detector, which travel downwards and exit.
The downward-going sample is consistent with the atmospheric-neutrino
flux prediction, but the upward-going sample exhibits a sizeable
depletion. Both are consistent with previously reported Soudan-2
neutrino-oscillation results. Inclusion of the two samples in an all-event
likelihood analysis, using recent 3D-atmospheric-neutrino-flux calculations,
reduces both the allowed oscillation parameter region and the probability
of the no-oscillation hypothesis. 


\end{abstract} 
\eject 
\large

\section{Introduction} 

   The  recently published neutrino oscillation analysis of the
Soudan-2 atmospheric-neutrino data~\cite{msanchez} used only
Fully Contained (FC) events and Partially Contained (PC) events
in which the neutrino interaction vertex was contained within the
Soudan 2 tracking calorimeter. Not utilized by that study were
two additional, topologically very similar, categories
for which further analysis was needed to separate the samples and to
eliminate their non-neutrino background. Using nomenclature introduced
by the MACRO experiment~\cite{MACRO-USID}, the two categories are
 labelled as {\it UpStop} events and {\it InDown} events.

{\it UpStop} events are upward-going stopping muons which arise from
charged-current $\nu_\mu$~and \anumu~interactions occurring in the rock
surrounding the Soudan 2 cavern; only the final state muon is detected
as a stopping, non-interacting track in the detector. The muon may be
accompanied by a small number of hits from a decay electron near its
stopping point.

{\it InDown} events are  $\nu_\mu$~and \anumu~interactions in the detector
yielding downward going, exiting muons with three or fewer hits 
arising from hadronic track(s) at the production vertex. 
Approximately 65\% of these InDown events 
are quasi-elastic interactions
with low energy protons.  Interactions having more than three
hadronic hits at their primary vertex were
included in the PC sample analyzed previously \cite{msanchez}.

Separation of these two neutrino event samples is made possible
by the fine-grain imaging of the 
Soudan-2 honeycomb-lattice tracking calorimeter. Their angular distributions
exhibit features which are indicative of atmospheric 
$\numu \rightarrow \nutau$ oscillations.  The samples can be 
incorporated in a straightforward way into the likelihood
analysis described in Ref. \cite{msanchez}. Their inclusion
 has enabled an improved determination of the 
$\numu \rightarrow \nutau $ oscillation parameters from this experiment
and a more stringent rejection of the no oscillation hypothesis.

Analyses of upward through-going muons
\cite{MACRO-USID, Kam-UT, IMB-USUT, SK-UT}, and
stopping muons \cite{MACRO-USID, IMB-USUT, Frejus-USID, SK-USUT}
have been previously reported. Whereas through-going muon
samples originate from a broad high-energy neutrino spectrum
having a mean \enu~ of approximately 100 GeV, UpStop events originate
predominantly from interactions with $1 \le \enu \le 20$~GeV.
Consequently they provide different constraints for oscillation 
scenarios.   Among the underground tracking calorimeter experiments, 
MACRO has provided the most detailed treatment to date of 
UpStop and InDown events.  In that experiment it was not possible 
to separate the two categories, so they were analyzed as a combined
sample.  Clearly, it is advantageous to separate the samples,
since comparison of their zenith angle distributions can provide
additional discrimination between low ($\approx 10^{-3}$ eV$^2$) 
versus high ($\ge 10^{-2}$ eV$^2$) values of \delmns.

\section{Detector and data exposure} 

Soudan 2 was a 963 metric ton (770 tons fiducial) iron tracking calorimeter
with a honeycomb geometry which operated as a time-projection chamber. 
The detector was located at a depth of 2070 meters--water--equivalent on the
27th level of the Soudan Underground Mine State Park in northern Minnesota.
The calorimeter started data taking in April 1989 and ceased operation in
June 2001 by which time a total exposure (fiducial exposure)
of 7.36 kton-years (5.90 kton-years) had been obtained.

The calorimeter's tracking elements were 1 m long, 1.5 cm diameter hytrel
plastic drift tubes filled with an argon-CO$_2$ gas mixture. The tubes
were encased in a honeycomb matrix of 1.6~mm thick corrugated steel plates.
Electrons deposited in the gas by the passage of charged particles drifted
to the tube ends under the influence of an electric field. At the tube ends
the electrons were amplified by vertical anode wires which read out a column
of tubes. A horizontal cathode strip read out the induced charge and the
third coordinate was provided by the drift time. The ionization deposited
was measured by the anode pulse height. The calorimeter produced
three-dimensional track hits with a spatial resolution of approximately
1 cm$^3$ and separated by an average of about 3 mm of steel.   
The corrugated plates, interleaved with drift tubes, were 
stacked to form 1$\times$1$\times$2.5 m$^3$, 4.3-ton modules
from which the calorimeter was assembled 
in building-block fashion~\cite{S2:NIM_A376_A381}. 

Surrounding the tracking calorimeter on all sides, but mounted on the cavern
surfaces and well separated from the outer surfaces of the calorimeter, was 
a 1700 m$^2$ Veto Shield array of two or three layers of proportional
tubes~\cite{S2:NIM_A276}.  The shield recorded the presence of cosmic
ray muons coincident in time with events in the main calorimeter 
and thus identified background
events, either produced directly by the muons or initiated by secondary
particles coming from muon interactions in the rock walls of the cavern. 
Additionally, for neutrino-induced muons which enter or exit the tracking
calorimeter, the shield array recorded the muon in time-coincidence with
the event in the central detector.  

\section{Separation of UpStop and InDown Events}
\label{sec:attrib}

 The event imaging afforded by the Soudan 2 tracking calorimeter
made it possible to distinguish the topologically similar UpStop
and InDown events. Events of both types were isolated during routine
processing.

Events were classified as UpStop candidates 
if they satisfied the following criteria:
\begin{enumerate}
\item [(1)] The track was muon-like, devoid of kinks or
       scatter vertices.
\item [(2)] The track length was greater than 100 cm.
\item [(3)] The muon endpoint occured in a live detector region. An event
       was removed if the candidate endpoint occured in the inactive
       region between modules. 
\item [(4)] Track ionization and straggling were consistent with the 
       hypothesis of an upward-going muon which ranges to stopping.
       That is, near the edge of the detector the track was straight and
       lightly ionizing while, near the interior end, the track exhibited
       multiple scattering and/or heavy ionization.
\item [(5)] Associated hits at the track endpoint, if any,
       had to be consistent with an electron shower
       from muon decay.
\end{enumerate}

\begin{figure}[htb]
\centerline{\epsfig{figure=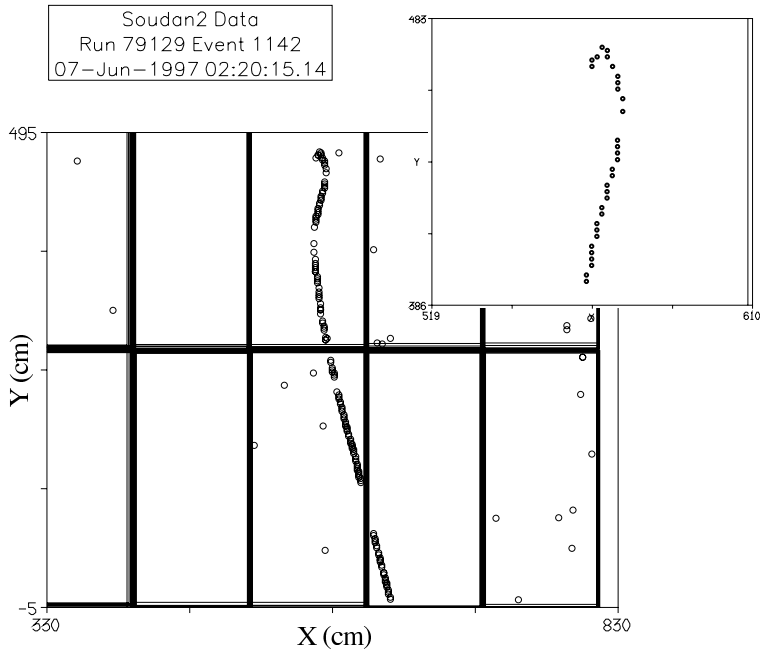,width=10.0cm}}
\caption{An UpStop data event recorded in the anode-cathode matched view
(front view of the calorimeter). Typically, multiple scattering becomes
pronounced as the muon approaches its range endpoint. 
The endpoint decay shower of three hits favors identification 
as a \mup rather than a \mum.}
\label{fig:fig1}
\end{figure}

   An anode-versus-cathode view of an UpStop data event 
is shown in Fig.~\ref{fig:fig1} where multiple scattering
can be discerned as the muon ranges to stopping.
Endpoint decay hits, the three hits modestly displaced
from the muon endpoint in Fig.~\ref{fig:fig1}, are observed in some events
(with higher probability for $\mu^+$ than for $\mu^-$
since the former do not undergo nuclear absorption within iron nuclei). 

  InDown events were required to satisfy criteria 1-3 above and in addition:
\begin{enumerate}
\item [(4)] The muon track was straight and lightly ionizing at its interior end.
Near the edge of the detector the track might, but need not,
have exhibited ranging behavior in the form of 
multiple scattering and/or heavy ionization. 
\item [(5)] Associated hits near the interior
end of the track, if present, must have been consistent  
with hits from a proton or $\pi^\pm$ track, lying
in a straight line and heavily ionizing.
\end{enumerate}
 
\begin{figure}[!htb]
\centerline{\epsfig{file=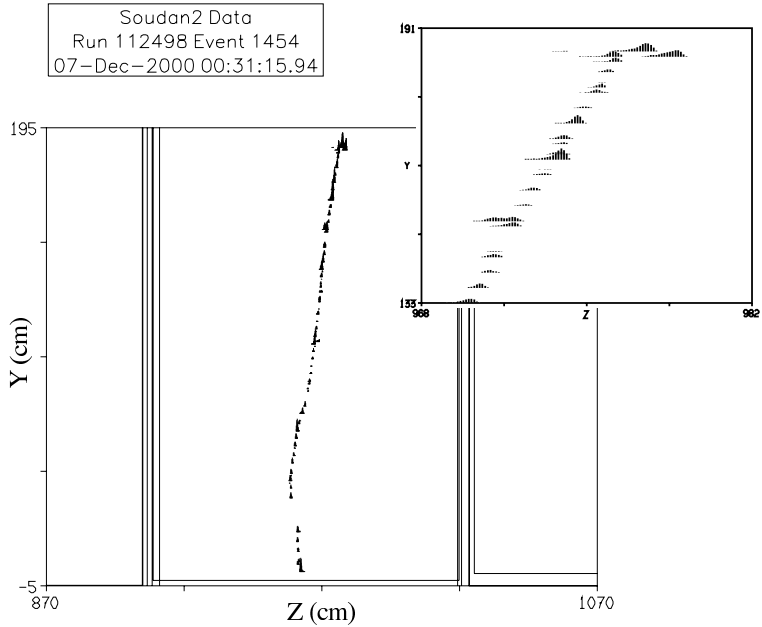,width=10.0cm}}
\caption{An InDown data event recorded in the cathode versus time view
(calorimeter side view). The muon emerges from 
 an event vertex which is well contained; its
 trajectory, a straight line initially, undergoes small angle deflection
 as the ranging muon approaches the detector floor.  A proton recoil of two
 hits is visible at the vertex.}
\label{fig:fig2}
\end{figure}

These topological features are exhibited by the InDown data event
shown in Fig. \ref{fig:fig2}. At the event vertex, the muon is
accompanied by a track of two hits for which the ionization is
relatively heavy. This pattern is typical of a recoil proton.
The event of Fig. \ref{fig:fig2} is a candidate quasi-elastic
$\numu n \rightarrow \mum p$.

There were a few events which could not be resolved as UpStop or InDown,
the direction of the track being undetermined.  Fortunately, ambiguous
cases were rare for tracks which have visible lengths exceeding one meter
in the detector. For the purpose of analysis, such events were retained
as an {\it Ambiguous} category.

\section{Event processing and simulation}
\label{section:sim}

Both UpStop and InDown data events were selected in the Partially Contained
event sample, and were processed as described in Ref. \cite{msanchez}.
The Monte Carlo sample of the contained-vertex InDown events was also part 
of the routine data processing, in which Monte-Carlo events were inserted 
into  and processed together with the data stream, their identity only being
revealed in the final analysis stage. However, additional simulations, not
included in the main data processing, were needed for the neutrino
interactions in the rock surrounding
the cavern, which give rise to UpStop events \cite{PDK-803}.

\subsection{Simulation of upward-stopping muon events}

The GEANT Monte Carlo program together with modified
Soudan-2 software provided the UpStop simulation (UpStop-MC).
A total of 68.7 million neutrino interactions in greenstone rock
were simulated. The event vertices were distributed randomly through
rock volumes which were centered on the Soudan 2 cavern. Since
high-energy charged-current (CC) events can project muons
to the cavern from more remote rock than low-energy  events,
the dimensions of the primary rock volumes were chosen to increase with 
increasing $E_{\nu}$. Final-state particles were tracked 
through the rock by GEANT and fourvectors of particles that reached 
the veto-shield array were saved.  These were then passed through the
Soudan-2 Monte Carlo to produce realistic detector hits superimposed
on detector noise represented by random-trigger records.

UpStop-MC events were then processed through the standard Soudan-2
triggering, reconstruction and selection software for PC events.
For UpStop-MC events yielding ionization within the tracking calorimeter
(71,000 events), the survival rates decreased  monotonically with increasing
primary \enu, reflecting the diminishing probability for muons from
energetic events to stop in the detector.  Survival rates ranged
from 12.8\% for events with \enu\ $\leq$ 10 GeV, to 2.5\% for events
with \enu\ $\geq$ 40 GeV.

\subsection{UpStop cuts and scanning}
\label{sec:cuts}

The PC selection filter required candidate tracks to penetrate to the
fiducial region while not being through-going. Additional requirements,
detailed in Ref. \cite{msanchez}, were imposed in order to reject the
high-flux background of downward-going cosmic-ray muons. A total of
7662 UpStop-MC events  passed the filter and simulated trigger requirements.
However, only 34\% of these events yielded a potentially
interesting topology in the detector. 
Consequently, additional cuts  were applied to the true kinematic variables 
to reject those events which were certain not to pass the subsequent
analysis cuts.
These cuts (existence of a final-state muon with cosine zenith angle, 
cos $\theta_z < +0.05$, and energy $E_\mu$ upon arrival at the detector
within the range $350~{\rm MeV} < E_\mu < 3500$ MeV)
were designed to ensure that the event had an upward-going muon 
that stopped within the calorimeter fiducial volume \cite{PDK-803}.
A 54\% sample of the surviving events was then scanned by physicists,
using scanning rules identical to those used for PC data event scanning. 
The additional criteria given in Sec. \ref{sec:attrib} were also applied 
to both data and MC events. All events which satisfied the scanning
criteria were then reconstructed manually using the experiment's standard
interactive graphics software.  Only the reconstructed sample was used
in the subsequent analysis.  It contained a factor of 25 more events
than the data sample. 

\subsection{MC event rate normalization}
\label{sect:norm}

The atmospheric neutrino flux used to generate the UpStop events was the
one-dimensional calculation of the Bartol group \cite{bartol89},
modulated by the solar cycle as described in Ref. \cite{msanchez}.
Other fluxes were simulated by applying correctional weights to the
generated events. For consistency with Ref. \cite{msanchez}, the
numbers and plots in Sects. \ref{sec:bgr} and \ref{sec:dist} were
weighted to correspond to the updated 1D Bartol-96 flux \cite{bartol96}.
The oscillation analysis of Sect. \ref{sec:global} of this paper
used the latest three-dimensional fluxes 
from the Bartol group \cite{Bartol04} and
 Battistoni {\it et al.} \cite{Battistoni_3D}.
The neutrino cross sections were those encoded in NEUGEN3 \cite{neugen}.
The target nuclear composition was that of Soudan rock, described in
Ref.~\cite{demuth}. The effect of Pauli blocking in elastic and
quasi-elastic reactions was accounted for, however nuclear effects
on resonance production and on deep inelastic scattering final states
were neglected.

The event rate calculations have a sizable systematic error.
For the comparison of this data with the MC
presented in Sections \ref{sec:bgr} and \ref{sec:dist}, a normalization
factor of 0.85, determined from the measured
$\nu_e$ rate, assuming no oscillations, was applied.  
In the oscillation analysis described in
Sec. \ref{sec:global} the overall normalization was a free ``nuisance''
parameter.

\section{Event rates and backgrounds}
\label{sec:bgr}

\subsection{Backgrounds in UpStop events}

Two sources of background events were considered:
\begin{enumerate}
\item [(1)] Cosmic-ray muons which scatter 
 in the rock and eventually enter
 the detector in an upward direction.
\item [(2)] Charged hadronic tracks, especially pions, produced
 at large angles in interactions
 of cosmic-ray muons in the rock surrounding the detector.
\end{enumerate}

Unlike experiments which are situated under mountains, the flat
overburden at Soudan ensures that the flux of cosmic-ray muons
becomes less than the flux of neutrino-produced muons significantly
above horizontal angles \cite{demuth}. Thus the background from
the first source is negligible. 

Hadronic interactions of neutrons produced in cosmic ray interactions
were shown in Ref. \cite{msanchez} to be a background
to contained neutrino events.  
A similar, related, flux of charged hadrons also emerges from the 
cavern walls and enters the detector.

  There are two distinguishing features of entering hadronic tracks:
\begin{enumerate}
\item [(1)] {\it Veto-Shield signal:} In addition to
 the Veto Shield hit corresponding to the passage of the hadron
track, there are likely to be extra hits due to other
particles produced in the muon interaction. 
In general, the track in the detector will not be aligned with
these extra hits. It is thus useful to distinguish between the total
number of in-time Veto Shield hits ($n^{VS}_{all}$) and the
number of in-time hits geometrically associated with the
incoming/leaving track ($n^{VS}_{trk}$).
\item [(2)] {\it Range:} Hadronic tracks have a limited range in
the detector before stopping or interacting.
\end{enumerate}

\begin{figure}[htb]
\centerline{\epsfig{figure=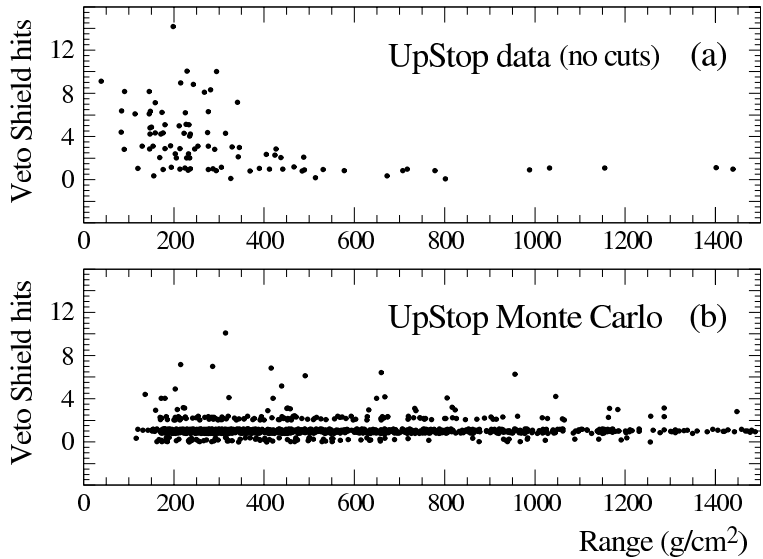,width=10.0cm}}
\caption{Distributions of Veto-Shield hit counts versus track range, (a)
for UpStop data candidates and (b) for UpStop Monte Carlo events,
prior to muon length cuts.}
\label{fig:fig3}
\end{figure}

\begin{figure}[htb]
\centerline{\epsfig{figure=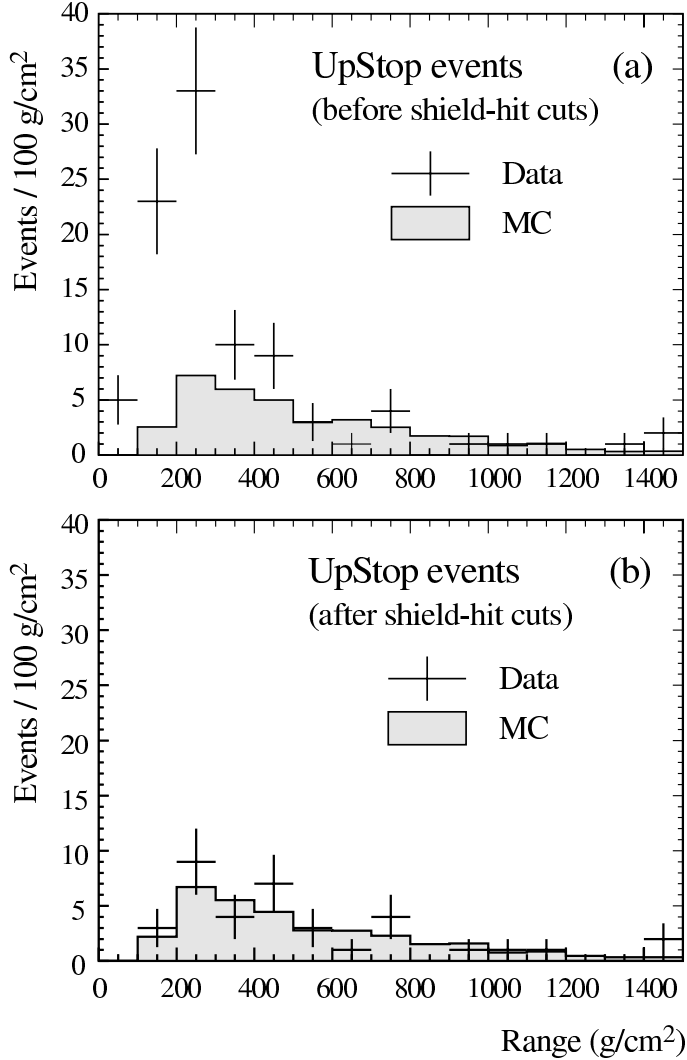,width=9.0cm}}
\caption{Distribution of track range for candidate UpStop data
events (crosses) compared to the neutrino Monte Carlo sample
(shaded histogram), for events with visible track length
exceeding one meter. 
The distributions (normalized to the events with $>$ 500 g/cm$^2$)
are shown (a) before and
(b) after the requirement $n^{VS}_{all} = n^{VS}_{trk}$ has been
applied.  Fig. \ref{fig:fig4}b shows good agreement between the
data and the neutrino MC.}
\label{fig:fig4}
\end{figure}


Fig. \ref{fig:fig3} shows $n^{VS}_{all}$ versus the range
of the stopping track for each event of the UpStop data
(Fig. \ref{fig:fig3}a) and of the MC (Fig. \ref{fig:fig3}b). A clear
excess of data events with large $n^{VS}_{all}$ and small
range is observed, corresponding to the expectation for
incoming hadronic tracks.  However, at track lengths beyond 
the hadronic range, the data is consistent with
the UpStop MC.
The background signature is emphasized in
Fig. \ref{fig:fig4}a which shows the projection onto the
range axis of the UpStop data (points with errors) and the MC
(shaded histogram). The MC is normalized to the data with
range $>$ 500 g/cm$^2$ ($>$ 3.8 pion interaction lengths).  
Fig. \ref{fig:fig4}b shows the
same distributions but with the additional constraint that
$n^{VS}_{all} = n^{VS}_{trk}$, i.e. all Veto-Shield hits must be
geometrically associated with the incoming track. The MC is
then in good agreement with the data.  However,
to ensure that the residual hadronic background is negligible,
a cut requiring the track range to be
greater than 260 g/cm$^2$, corresponding to two pion interaction
lengths, is also applied.  Since the calorimeter is, to good approximation,
a uniform medium of 1.6 g/cm$^3$ density, the range requirement 
corresponds to a minimum track length requirement of $\sim$160 cm.
The effective muon momentum threshold for UpStop and InDown events is
$p_\mu \geq 530$ MeV/$c$.

Finally, the cosine of the zenith angle, cos$\theta_z$, of the
reconstructed UpStop track is required to be smaller than +0.05.

\subsection{UpStop backgrounds using hadronic scatter events}

\begin{figure}[htb]
\centerline{\epsfig{figure=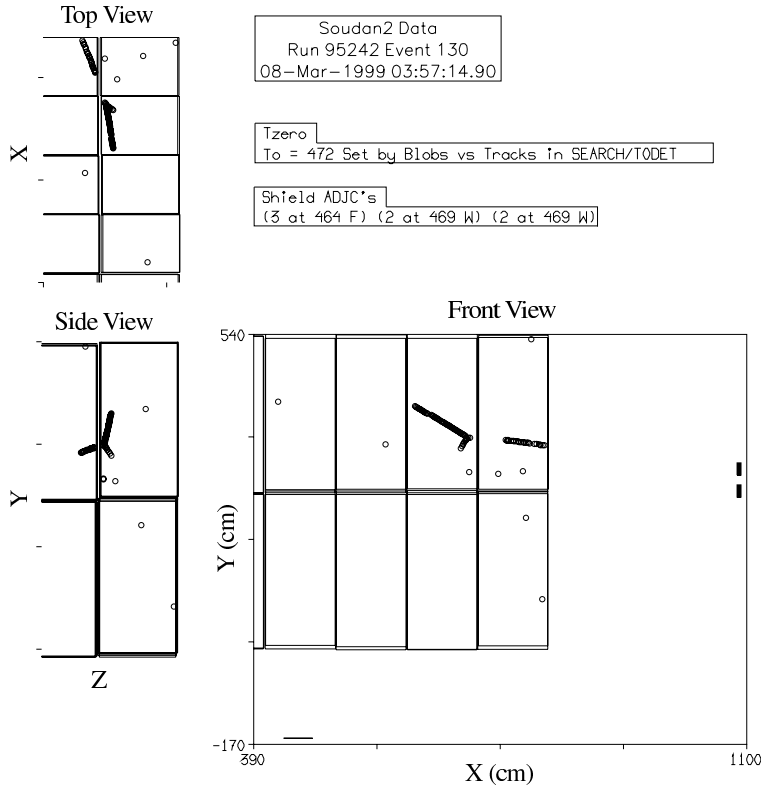,width=10.0cm} }
\caption{A scattering upgoing pion track (data event) accompanied by three
time-coincident Veto Shield hits.}
\label{fig:hadevt}
\end{figure}

The PC data analysis also recorded events with an incoming
track making a hadronic scatter.  A sample of 25 data events was
obtained which can be used to gauge the background from incoming
non-scattering hadrons. A  representative event is the upward-going,
stopping, charged pion track shown in Fig. \ref{fig:hadevt}.
There are two coincident hits in the shield which are in close
proximity to the track's entrance point into the cavern, hence
$n^{VS}_{trk} = 2$.  There is an additional coincident hit in
the shield floor, hence $n^{VS}_{all} = 3$. Inside the detector,
the track scatters and stops. The scatter, which gives rise
to a recoil proton,  plus the track ionization, identifies the track
as a $\pi^{\pm}$ rather than a $\mu^{\pm}$ or a proton.

\begin{figure}[htb]
\centerline{\epsfig{figure=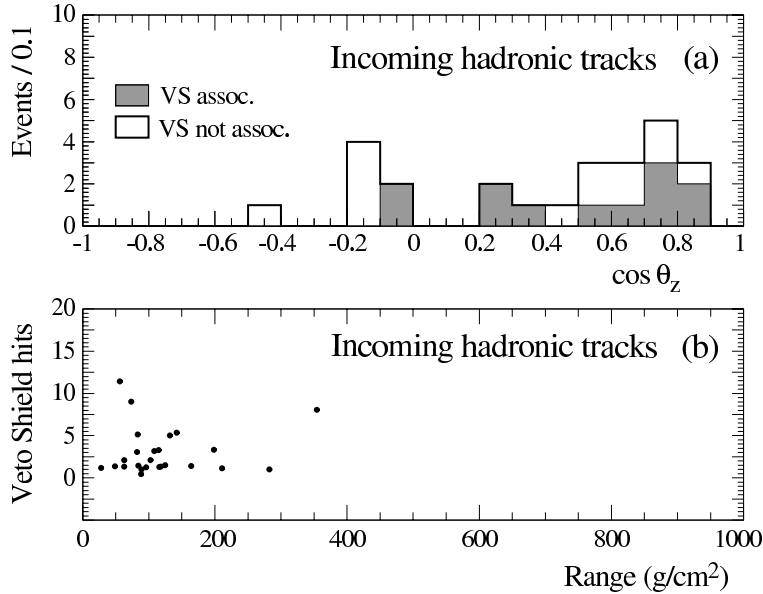,width=10.0cm}}
\caption{Distributions of (a) cos$\theta_z$ and
 (b) the number of coincident veto shield hits versus the
 track range, for events which exhibit hadronic scatters.
 In Fig. \ref{fig:had_3}(a), the shaded (open) histogram includes
 events where all the in-time Veto-Shield hits
 are (are not) associated with the detected track. }
\label{fig:had_3}
\end{figure}

Distributions of cos$\theta_z$ and of track range versus $n^{VS}_{all}$
for the incoming hadronic sample are shown in Fig. \ref{fig:had_3}.
Fig. \ref{fig:had_3}a shows that most of the events are downgoing.
Among the seven upward-going tracks, five have
$n^{VS}_{all} > n^{VS}_{trk}$ (open histogram); the remaining two events
with $n^{VS}_{all} =  n^{VS}_{trk}$  (shaded histogram) are close to
horizontal, and have short range. No hadronic events pass the shield and
track length selections for upgoing tracks (ignoring their visible scatters).  
On the basis of this observation, plus the agreement in track range
distributions between the UpStop data and the neutrino MC of
Fig. \ref{fig:fig4}, backgrounds from non-scattering charged pions
and protons are estimated to contribute less than two events to the UpStop
data  and are hereafter neglected.

\subsection{Backgrounds in InDown events}
 
A potential background for InDown events arises from
downward, through-going cosmic ray muons
whose entrance into the detector was not detected due to a rare episode 
of poor or non-existent ionization drifting within a calorimeter module.
Great care was taken to record all such incidents during data taking 
and additional checks were made by studying individual
module efficiencies as a function of time. A special scan was carried
out which rejected events for which there was a possibility of
such an occurrence.  Additional discrimination against this background was
provided by the active shield array, since through-going muons yielded pairs
of time-coincident hits having widely separated spatial locations.
 Using the same minimum track range as was used for the UpStop sample
($\ge$ 260 g/cm$^{2}$), no InDown events had $n^{VS}_{all} > n^{VS}_{trk}$.
Other backgrounds for PC events were shown to be negligible in
Ref. \cite{msanchez}, thus the InDown sample was assumed to be
background-free in the analysis presented below.

\subsection{Event rates}

A final sample of 1081 fully reconstructed UpStop-MC events was retained
for subsequent analysis.  Within this simulation sample, 80\% of events
originate with neutrino interactions having \enu~$\leq$ 10 GeV.
Charged-current quasi-elastic scattering 
accounts for about one third of the interactions.
The neutrino fraction \numu /(\numu + \anumu) of the sample is 64\%.

The numbers of candidate data neutrino events which survive are listed in
Table \ref{tbl:counts}, where the MC numbers have been scaled to an
exposure of 5.90 fiducial kton-years and include the factor of 0.85 
to normalize to the \nue event rate of Ref. \cite{msanchez}.  
Comparison of the data with the neutrino MC
predictions, the sum of columns 2 and 3, shows that the observed InDown
rate is consistent with the prediction, whereas the UpStop data rate appears
suppressed by a factor of approximately two. These trends are in agreement
with the expectation from the oscillation analysis of Ref. \cite{msanchez}.
Note also the small size of the ambiguous sample and the small
misidentification rate between the UpStop and InDown samples.

\begin{table}[hbt]
\centering
\caption{Numbers of data and Monte Carlo events which pass all cuts.
The no-oscillation MC event rate is normalized to the measured e-flavor
event rate of Ref. \cite{msanchez} assuming no oscillations.}
\begin{tabular}{|l|c|c|c|}
\hline
 {\it Assigned as} & \multicolumn{2}{c|}{\it No-osc. MC Truth} &
 \emph{Data}\\ \cline{2-3}
     & InDown & UpStop & \\
\hline
InDown  & 12.4$\pm$1.4 & 0.3$\pm$0.1 & 16 \\
UpStop    & 1.8$\pm$0.5  & 53.3$\pm$1.8 & 26 \\
Ambig   & 0.8$\pm$0.3 & 3.4$\pm$0.4 & 2 \\
\hline
\end{tabular}
\label{tbl:counts}
\end{table}

\section{Energy and angular distributions of UpStop/InDown neutrinos}
\label{sec:dist}

\begin{figure}[hbt]
\centerline{\epsfig{figure=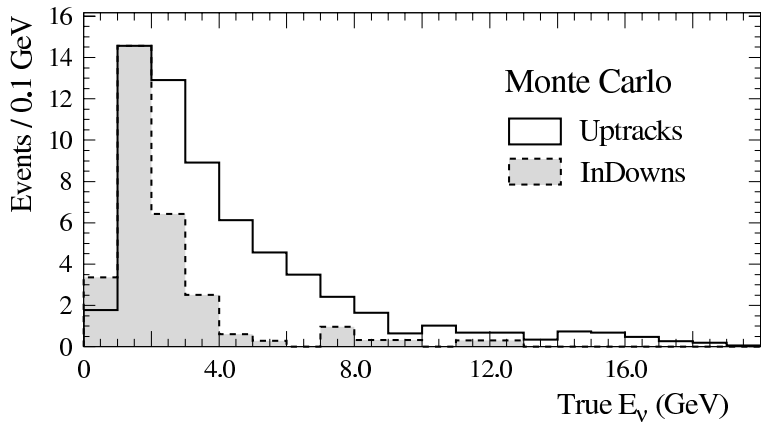,width=10.0cm}}
\caption{Comparison of the primary $E_\nu$ spectra for UpStop and InDown
 events.}
\label{fig:truen}
\end{figure}

The neutrino energy, $E_\nu$, for Monte Carlo UpStop and InDown events 
is shown in Fig. \ref{fig:truen}. The UpStop events have an average
$E_\nu$ of 6.2 GeV. In contrast, the InDown events have lower $E_\nu$
values with an average of 2.4 GeV.

The muon track provides a good  estimator for the incident neutrino
direction. For UpStop events the average angle between the incoming
neutrino and the muon is 11$^\circ$. For the lower energy InDown sample,
the average angle is 13$^{\circ}$.

\begin{figure}[hbt]
\centerline{\epsfig{figure=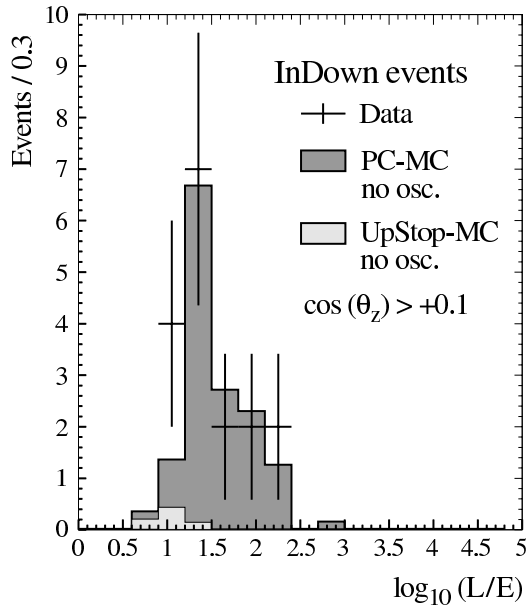,height=8.0cm}}
\caption{Distribution of log$_{10}(L/E)$ for the InDown events. 
The data are shown as crosses; the solid-line histogram shows the
neutrino MC expectation for no oscillations. The MC prediction
arises mostly from contained vertex PC events (PC-MC, dark shading),
but receives a small contribution from UpStop-MC events which were
misidentified as InDowns (light shading).
}
\label{fig:ind}
\end{figure}

In the likelihood analysis of Ref. \cite{msanchez}, the variable
log$_{10}(L/E)$ was used, where $L$ is the neutrino pathlength in
kilometers and $E$ is the neutrino final state visible energy in GeV.
As demonstrated in Ref. \cite{msanchez}, multiple scattering of the
exiting muon in PC events allows sufficient neutrino energy determination
to permit the reconstruction of the log$_{10}(L/E)$ distribution. 
Fig. \ref{fig:ind} shows the log$_{10}(L/E)$
distribution for InDown data (crosses) compared to the no-oscillation MC.
The solid-line histogram represents the sum of the partially-contained
InDown events from the contained-vertex PC Monte Carlo (PC-MC, dark
shading), plus misidentified UpStop-MC events (light shading). 
Good agreement is observed between the data and the unoscillated Monte
Carlo. Hence no evidence is found for oscillations in $\nu_\mu$ atmospheric
neutrinos which are incident from directions above the horizon.

\begin{figure}[hbt]
\centerline{\epsfig{figure=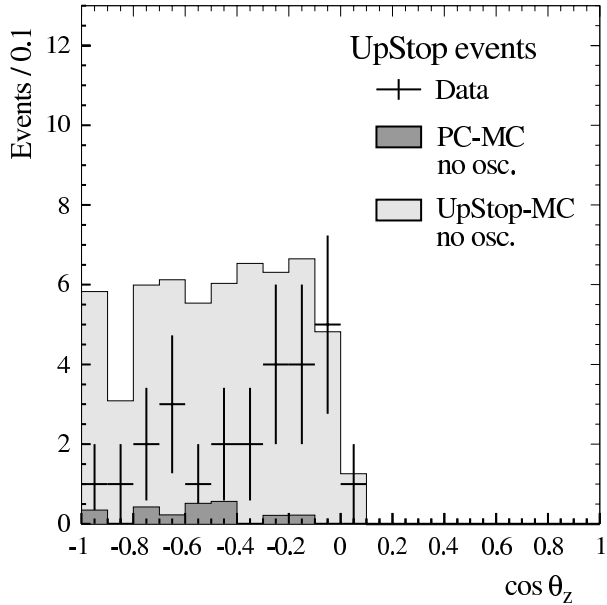,height=8.0cm}}
\caption{Distribution of cos$\theta_z$ for the UpStop events.
The crosses represent the data, and  the solid-line histogram shows the
expected MC distribution for no oscillations.   Light and dark-shaded
areas show the contributions of UpStop and InDown events
for no oscillations.}
\label{fig:upt}
\end{figure}

For the UpStop events, however, the whole of the hadronic-shower energy
plus a fraction of the muon energy is missing.  Thus, whereas $L$ can
be calculated accurately from $cos\theta_z$, $E$ is essentially
undetermined. The average difference between log$_{10}(L/E)_\nu$
and  log$_{10}(L/E)_\mu$ is 0.69 (FWHM=0.66),
spanning four bins used in the oscillation analysis.  Therefore
log$_{10}(L/E)$ is not a useful variable for analysis of oscillations
in these events. In this section we show data as a function of
cos$\theta_z$. In the analysis described in Section \ref{sec:global}
it is convenient to fit as a function of log$_{10}L$. Fig. \ref{fig:upt}
shows the distribution of cos$\theta_z$ for upward-stopping muon data
events. The distribution decreases steadily towards the nadir.
Fig. \ref{fig:upt} also includes the  simulated
 UpStop-MC (light shading) and the misidentified
contained-vertex PC-MC (dark shading) distributions, for the no-oscillation
case. Significant disagreement between the neutrino UpStop events and
the no-oscillation expectation is apparent towards the nadir, 
which is consistent with the loss of 
upward-going $\mu$-flavor events due to oscillations.

  The distributions of Figs. \ref{fig:ind} and \ref{fig:upt}
imply constraints on neutrino oscillation scenarios.  
These samples have been included in a likelihood analysis together with all 
the other neutrino events from the experiment.  The method and the
results of this global fit to Soudan-2 neutrino data
are described in Section \ref{sec:global} below.

\section{Oscillation analysis}
\label{sec:global}

\subsection{Outline of the method}
\label{sec:method}

The oscillation analysis is a bin-free likelihood analysis based on
the prescription of Feldman and Cousins \cite{Feldman_Cousins}.
A detailed description of the method can be found in Ref. \cite{msanchez}
and is not repeated here. The main points of the analysis are:

\begin{enumerate}
\item [(1)]  As reported in Ref. \cite{SK-USUT} and confirmed by
Ref. \cite{msanchez}, the distributions of neutrino-induced
e-flavor data events are consistent with the null oscillation
MC predictions, up to an overall normalization.
Only the $\mu$-flavor data  
exhibit oscillation effects. Thus this analysis
assumes two flavor \numutonutau oscillations.  

\item [(2)] The FC and PC samples described in Ref. \cite{msanchez}
are used unchanged in the present analysis. The InDown
events are added to the PC muon-flavor sample,
and the Ambiguous events are used only in the overall normalization.
The UpStop events are treated as a new category and analyzed
as a function of log$_{10}L$ rather than as 
a function of log$_{10}(L/E)$.

\item [(3)] A likelihood function for the data is constructed as a function
of \delmns, the mass-squared difference, and of \sqsinns, 
where $\theta$ is the mixing angle, 
using probability density functions (pdf's) determined from the MC sample.   
Details of the formalism are given in Ref. \cite{msanchez}.

\item [(4)] The summed negative log likelihood is evaluated at each point
on a 15 $\times$ 80 grid of \sqsin $\times$ log$_{10}(\Delta m^2)$ 
with \sqsin varied between 0.0 and 1.0 and
\delm varied between $10^{-5}$ and $10^0$ eV$^2$. The lowest negative
log likelihood on the grid is found and $\Delta \cal{L}$, the difference
between the lowest value and the value in each (\sqdm ) grid square, 
is plotted.

\item [(5)] A background contribution of non-neutrino events 
arising from neutrons and gammas produced by muon interactions 
in the rock around the detector is added to the likelihood
function.  The background contribution only affects the FC events; the
PC events and the new InDown and UpStop events are treated as
background-free.

\item [(6)] The overall normalization of the MC and the amounts
of background in the different FC event samples, estimated using
shield-tagged data events and the depth distribution of the event vertices, 
are nuisance parameters whose values are optimized at each grid square.

\item [(7)] The allowed confidence level regions are calculated by the
method of Feldman and Cousins \cite{Feldman_Cousins}. That is, MC
experiments are generated and analyzed at each grid square to
calculate the expected likelihood rise for a given confidence
level based on the statistical and systematic errors at that
grid square.  In addition to the systematic errors described
in Ref. \cite{msanchez}, a 10\% systematic error 
on the relative normalization of the UpStop events to the remainder 
of the data was allowed.  The latter error represents uncertainties in 
density and nuclear composition of rock below the detector, and uncertainties with
variation of neutrino cross sections in rock versus iron.

\item [(8)] The analysis of Ref. \cite{msanchez} used the one-dimensional
flux calculation of the Bartol group \cite{bartol96}.  
This analysis uses their new three-dimensional calculation \cite{Bartol04}
and compares it with their 1D calculation and with the 3D calculation 
of Battistoni {\it et al.} \cite{Battistoni_3D}. 
 
 \end{enumerate} 
 
 The 44 new data events documented here are added to
the 488 data events of the previous analysis.  However, the new events
are ``high resolution'' $\mu$-flavor events, those most sensitive
to oscillations, consequently they enhance the sensitivity afforded by the
 167 events of that type in the previous analysis.

\subsection{Oscillation results}
\label{oscanal}

\begin{figure}[htb]
\centerline{\epsfig{file=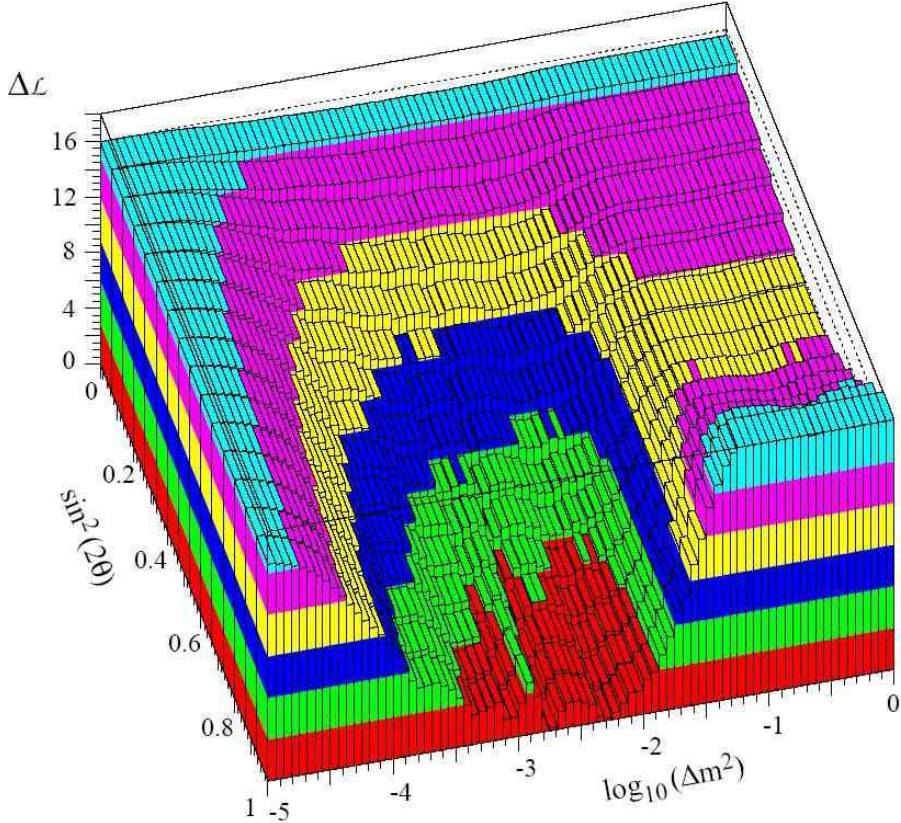,width=12.0cm}}
\caption{The data likelihood difference, $\Delta \cal{L}$, plotted as a
function of \sqsin and log$_{10}(\Delta m^2)$.}
\label{fig:datallh}
\end{figure}

The values of $\Delta \cal{L}$  are plotted 
in Fig. \ref{fig:datallh} as a function of \sqsin and log$_{10}(\Delta m^2)$.
The resulting surface is similar to that reported in Ref. \cite{msanchez}.
The main difference is that the likelihood rise for the grid square with the
lowest values of \delm and \sqsin (called
the no oscillation point) is 16.0 compared with 11.3 for
the previous analysis. The new data has  
significantly increased the discrimination against the
no-oscillation hypothesis, mostly because of the large
suppression of the UpStop data compared to the MC prediction.  
The probability of the validity 
of the no-oscillation hypothesis is discussed in Sect. \ref{sec:CL}.  

The  $\Delta \cal{L}$ surface in Fig. \ref{fig:datallh} exhibits
two nearly equal minima, one at the grid square centered at
\delm~=~0.0017~eV$^2$, \sqsin~=~0.97, and one at
 \delm~=~0.0052~eV$^2$, \sqsin~=~0.97.
The first minimum is the ``best fit point" of this
analysis, while the second minimum was the best fit point in the
previous analysis \cite{msanchez}.
The difference  of $\Delta \cal{L}$ between the two is only 0.18.
 There is a small rise in likelihood between the two minima which peaks 
 at about 1.8 in the region of the Super-K best-fit point. 
 However, the 90\%-confidence-level limit of this analysis, 
 determined in Sect. \ref{sec:CL}, 
 contains all of the Super-K allowed region. 
 The value of the flux normalization at the best fit point is 91\% of
 the Bartol-3D prediction~\cite{Bartol04}.

\begin{figure}[htb]
\centerline{\epsfig{file=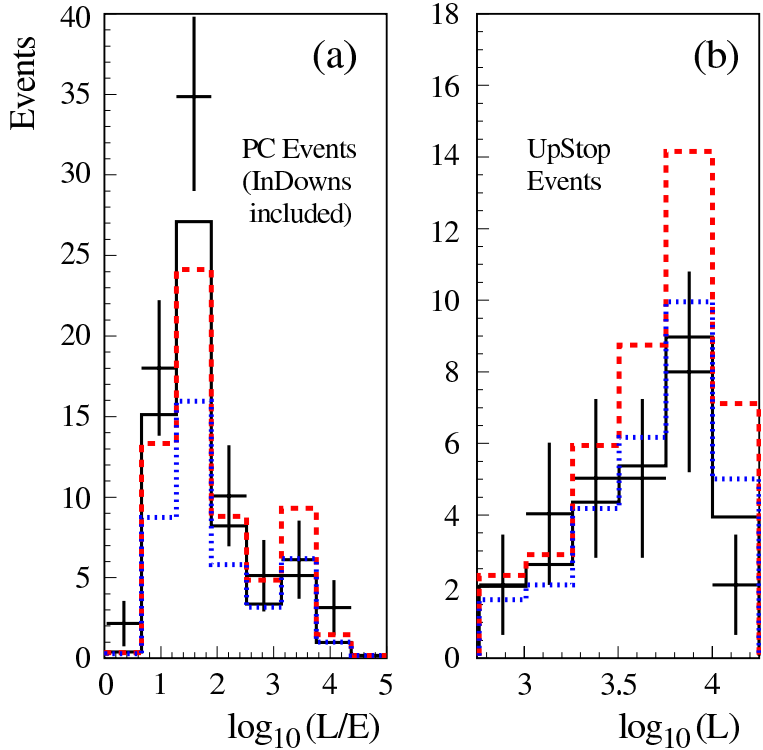,width=10.0cm}}
\caption{(a) The \ltenloe distribution for the PC event sample
that includes the InDown events.
(b) The UpStop events plotted as a function of log$_{10}L$.
The points with error bars are the data. The solid histogram
is the MC prediction at the best fit point, and the dashed
histogram shows the no-oscillation expectation.  The dotted histogram depicts
``saturated oscillations" with \sqsin~=~1.0, \delm~=~1.0~eV$^2$.}
\label{fig:inup}
\end{figure}

Although the analysis is carried out using the log-likelihood
function, it is useful to evaluate the goodness of the fit by
projecting out the distributions for the various data sets and
calculating a $\chi^2$ for the data compared to the MC prediction.
Fig. \ref{fig:inup}a shows the data for the total muon PC sample
(including the InDown events) and Fig. \ref{fig:inup}b shows 
the UpStop events, compared to
the MC predictions.   

The PC events are in good agreement
with the no-oscillation histogram, but disagree, 
particularly at low $L/E$ (downward going $\nu$ events), 
with the prediction at the highest considered values of
\delm and \sqsin ($\sim$1.0 eV$^2$ and $\sim$1.0), where
the oscillations have the greatest frequency and largest
size (called ``saturated oscillations''). 
On the other hand, the UpStop events are in disagreement 
with the no-oscillation prediction, particularly at large $L$, 
but agree with the saturated oscillation prediction.  
This behavior is indicative of oscillations with values 
of \delm in the low $10^{-3}$ eV$^2$ region where downward-going 
neutrinos have not yet oscillated  and  upward-going neutrinos
have saturated oscillations. The best-fit prediction agrees
well with both distributions. 
   
Table \ref{tbl:chisq} gives the $\chi^2$ for the comparison
of data and MC for the PC and UpStop distributions as well as
the full data set including the other distributions described
in Ref. \cite{msanchez}. The $\chi^2$ at the best fit point,
using all of the data, is 35.6 for 30 data points.

\begin{table}[hbt]
\centering
\caption{The $\chi^2$ for comparisons of the data to various
MC predictions for the PC events, the UpStop events, and
the full data set. 
}
\begin{tabular}{|l|c|c|c|}
\hline
  & \multicolumn{3}{c|}{$\chi^2$/Number of data points} \\
 \cline{2-4}
     & PC & UpStop &All data \\
\hline
Best fit   & 5.0/5  & 0.6/4 & 35.6/30 \\
No oscillations  & 5.9/5 & 7.9/4 & 66.6/30 \\
Saturated oscillations   & 19.8/5 & 1.9/4 & 63.2/30 \\
\hline
\end{tabular}
\label{tbl:chisq}
\end{table}

\subsection{Confidence levels}
\label{sec:CL}

\begin{figure}[htb]
\centerline{\epsfig{figure=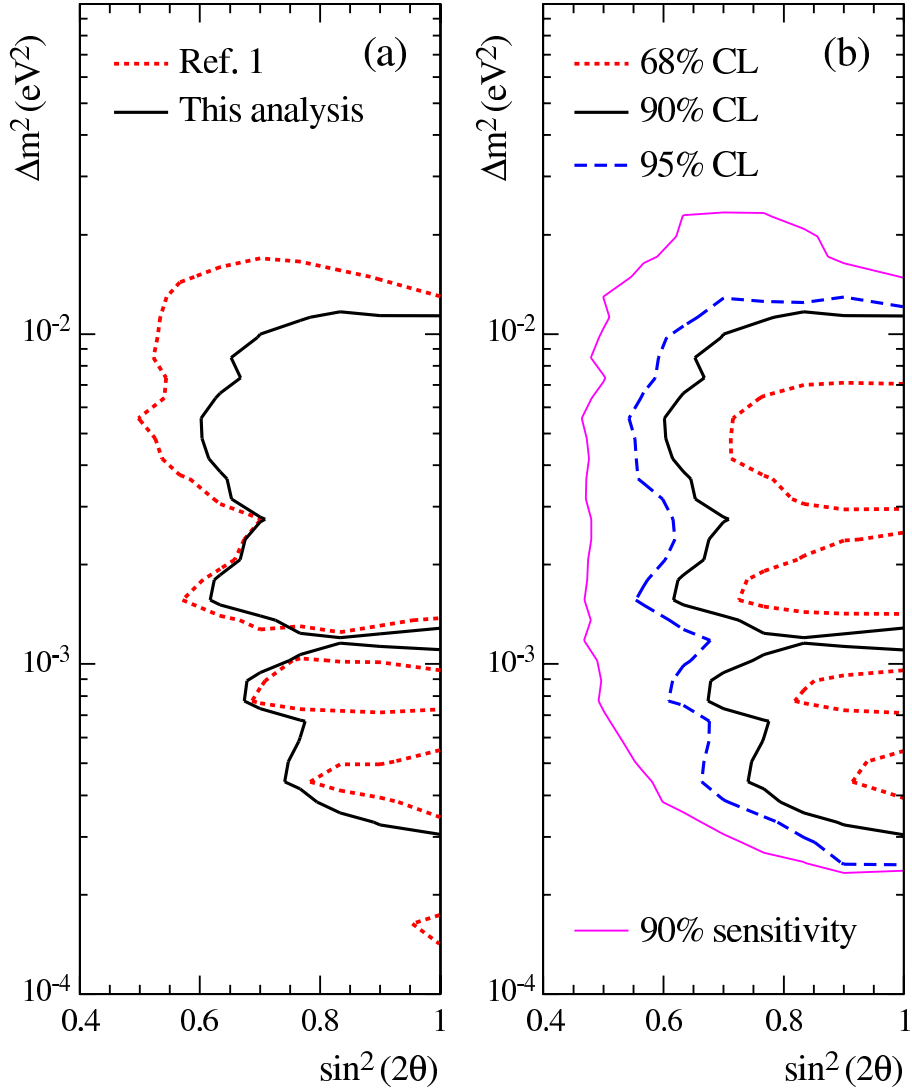,width=12.0cm}} 
\caption{(a) The 90\% confidence allowed region from the Feldman-Cousins
analysis of this work (solid line), compared to that of
 Ref. [1] \cite{msanchez} (dotted line). 
(b) Contours at 68\%, 90\%, and 95\% confidence level (dotted line,
thick solid line, and dashed line respectively), compared to 
the 90\% sensitivity contour (thin solid line).}
\label{fig:compnew}
\end{figure}

The 90\%-confidence-level surface $\Delta {\cal L}_{90}$ (not shown),
generated by the MC experiments
under Feldman-Cousins prescription is very similar to that of
the previous analysis.  Combining the 
$\Delta {\cal L}_{90}$ surface with the data
likelihood surface of Fig. \ref{fig:datallh} gives 
the solid-line 90\%-confidence-level contour
shown in Fig. \ref{fig:compnew}a. For comparison, the
90\%-confidence-level contour determined previously 
in Ref. \cite{msanchez} is shown by the dotted line.  It can be seen
that the allowed region is more restrictive and that some
of the contour structure indicated by the previous analysis has
been smoothed.  This is due to the fact that
the likelihood surface is rather flat at the base of the
valley and small changes in the data can move the contours
substantially in this region. 

Fig.\ref{fig:compnew}b shows the 68\%, 90\% and 95\% contours, 
a further indication of the shape of the likelihood surface. Also
displayed in Fig.\ref{fig:compnew}b is the 90\% sensitivity contour obtained 
from the Monte Carlo experiments, which denotes the expected 90\% contour 
for experiments with this statistical precision and systematic errors.  
As was the case in Ref. \cite{msanchez}, the 90\%-CL contour from this
analysis is more restrictive than the estimated sensitivity contour,
due, in part, to  a small mismatch of the overall event-rate
normalization in the electron and muon samples.    

The probability of no oscillations 
can be calculated, under the Feldman-Cousins prescription and
including all of the systematic effects, by generating experiments at the
minimum \delm and \sqsin grid square and counting those
MC experiments that give a larger likelihood difference
than 16.0.  In 500,000 simulated experiments, 
16 had a larger likelihood difference giving a probability of
$3.2\times10^{-5}$ for the no-oscillation hypothesis. 

  For UpStop muons and upward-going contained vertex events,
the initiating neutrino may experience Mikheev-Smirnov-Wolfenstein
(MSW) resonance and other matter effects as a result of the 
traversal of thousands of kilometers of terrestrial matter. 
The magnitude of matter-induced deviations from vacuum oscillations
was studied using simulated event samples; the samples were
weighted in accordance with three-neutrino mixing and the
normal mass hierarchy, using the approximation of a uniform
(path-weighted mean) terrestrial density \cite{G-Garcia_RMP03}.
For the range of plausible \delmns~ values, it was found that 
matter effects, even with maximally allowed mixing angles, can only
introduce a few percent additional depletion of muon-flavor neutrinos
beyond that which results from \numutonutau vacuum
oscillations \cite{PDK-811}. Since the scale of 
such deviations is well below the statistical sensitivity 
afforded by the data, matter effects were neglected
in this analysis.

\subsection{Flux model comparison and event rate normalization}
\label{sec:norm}
\begin{figure}[htb]
\centerline{\epsfig{file=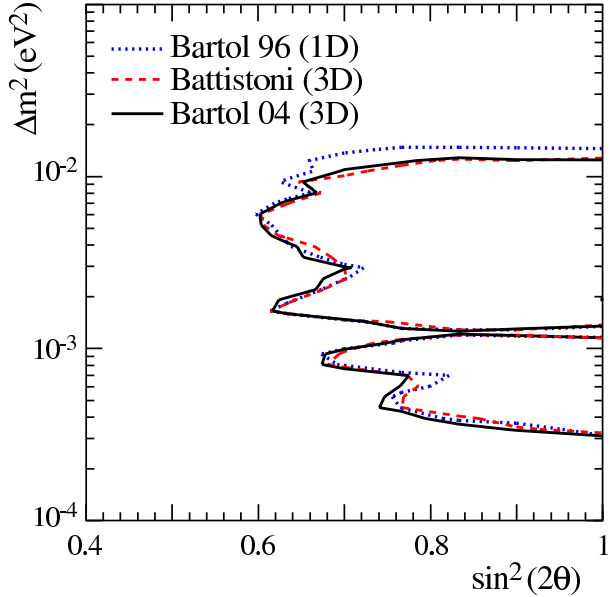,width=8.0cm}}
\caption{The 90\%-confidence-level allowed region for three
atmospheric-neutrino flux calculations.
The dotted curve is based upon the 1D-model
of the Bartol group from 1996 \cite{bartol96}.
The more recent 3D calculations of
Battistoni {\it et al.}  \cite{Battistoni_3D} and of
the Bartol group \cite{Bartol04}
lead to the dashed and solid curve, respectively.}
\label{fig:flux}
\end{figure}

The analysis of this data has been carried out 
for three different atmospheric neutrino flux calculations: {\it (i)}
the one-dimensional flux calculation 
of the Bartol group \cite{bartol96}, 
{\it (ii)} the Bartol three-dimensional
calculation \cite{Bartol04},
and  {\it (iii)} the three-dimensional calculation of
Battistoni {\it et al.} \cite{Battistoni_3D}.  
As well as the flux prediction, the analysis requires an estimate 
of the height in the atmosphere at which the neutrino is produced.  
This is particularly important for neutrinos coming from overhead 
where the path in the atmosphere is a large fraction of the total
path length. A parameterization of the pion and muon decay heights 
was made using the formalism of Ref. \cite{Ruddick} 
for the one-dimensional Bartol calculation. A similar
parameterization for decay heights  was prepared 
for the three-dimensional Bartol case. The latter parameterization
 was also used for predictions based upon the Battistoni {\it et al.} flux.

Fig. \ref{fig:flux} shows the 90\%-CL region for the three cases.  
There is a small change from the one- to three-dimensional flux models,
however the three-dimensional models of Bartol and Battistoni {\it et al.}
give almost identical results.
The only significant difference between the three cases 
is in the absolute normalization of the flux.  
At the best-fit point of this analysis, the normalization factor 
(number of events observed/calculated)
for the Bartol 1D flux is 0.86, while for the Bartol 3D flux the factor
is 0.91 and for the Battistoni {\it et al.} flux it is 1.02.

The authors of the flux calculations typically quote large errors of $\pm$20\%
on the absolute normalization, due to the uncertainties on the
incoming cosmic-ray fluxes and on nucleus-air cross sections.
There are also significant errors on the neutrino cross sections.  
It is thus of interest
to determine the experimental error on the 
ratio of the measured to the predicted event rate.
The experimental event rate is proportional to the incident neutrino flux,
the neutrino cross sections in the detector, and the detector acceptance.
This experiment can thus determine the normalization of the atmospheric
neutrino flux at the Soudan-2 site 
times the neutrino cross sections 
encoded in the NEUGEN3 program \cite{neugen}, 
for an iron calorimeter with a given energy threshold.  
Translation of this normalization factor to
other experiments at the Soudan site and at other sites is possible
in principle.  However it requires knowledge of the 
relative neutrino fluxes at the different sites and 
the ratio of the neutrino cross sections 
if a different detector medium or a different neutrino generator is used.

The event-rate normalization factor for Soudan 2 is subject 
to the following errors:
\begin{enumerate}

\item [(1)]  The total number of neutrino events observed in this experiment
above an energy threshold of 300 MeV, obtained from Table I of
Ref. \cite{msanchez} and Table 2 of this paper, is 481.2$\pm$26.2 ($\pm$5.4\%).
The error includes the statistical error and the error on the background
subtraction.

\item [(2)] The statistical error on the Monte Carlo sample is $\pm$1.6\%.

\item [(3)] The variation of the fitted normalization factor over the 68\% confidence
region of the oscillation parameters is $\pm$3\%.  The Feldman-Cousins 
analysis includes the systematic errors associated with the background
subtraction, cross section uncertainties and energy scale uncertainties.

\item [(4)] Depending on the value of $\theta_{23}$,
$\nu_e \rightarrow \nu_\mu$ oscillations with the solar parameters
could change the flux of $\nu_e$ that have traversed 
the Earth \cite{Peres-Smirnov, PDK-807}.  
The change can be positive or negative
depending on whether $\theta_{23}$ is smaller or greater than $45^o$.
Using the Super-K limits for $\sin^{2} 2\theta_{23}$ and recent values for
the solar oscillation parameters \cite{SK-USUT, Kamland}, 
an uncertainty of
$\pm$3.3\% in the calculated electron-event rate is estimated.

\item [(5)] Uncertainty arises in the rate of multi-GeV muon
events due to matter effects \cite{PDK-811}.  A $\pm$1.6\% error contribution
to the event rate calculation is inferred.

\item [(6)] Any mismatch between the Monte Carlo representation of
the detector and reality could introduce a relative error in the
acceptance of the two and thus an error in the normalization ratio.
Detailed studies of individual event channels revealed no significant 
differences \cite{Chung, PDK-791}. The relative proportions 
of different event types 
and event rejection modes in data versus Monte
Carlo samples were studied.  
A $\pm$2\% systematic error, estimated from
the maximum differences found between the data and Monte Carlo, has been
assigned to account for uncertainties arising from geometric acceptance  
and other residual effects. 
\end{enumerate}

Based upon the $\pm$8\% quadrature sum of these errors, 
an overall normalization factor of 0.91$\pm$0.07 
is determined for this analysis.  
This normalization is specific to the Soudan 2 site,
the detector medium, the Bartol 3D flux, and to the neutrino cross sections
encoded in the NEUGEN3 event generator.
It is averaged over the years 1989-2001, one full solar cycle.
The same percentage error, $\pm$8\%, is applicable 
to the Soudan-2 normalization
factors given above 
for the Bartol 1D and Battistoni {\it et al.} 3D
atmospheric fluxes.

As a check, a normalization which is mostly independent of
the $\nu_\mu \rightarrow \nu_\tau$ oscillation parameters can
be obtained from the total electron sample of contained and
partially contained events.  Table I of Ref. \cite{msanchez}
lists 208.7$\pm$15.9 background-subtracted electron-neutrino
events to be compared with an expected rate from the 3D Bartol
prediction of 238.1 events, yielding a normalization factor of
0.88$\pm$0.07, where the error is just statistical from the number
of events and does not include the contribution from
$\nu_e \rightarrow \nu_\mu$ oscillations or the other error
sources detailed above.

\section{Conclusions}
  
\begin{figure}[htb]
\centerline{\epsfig{file=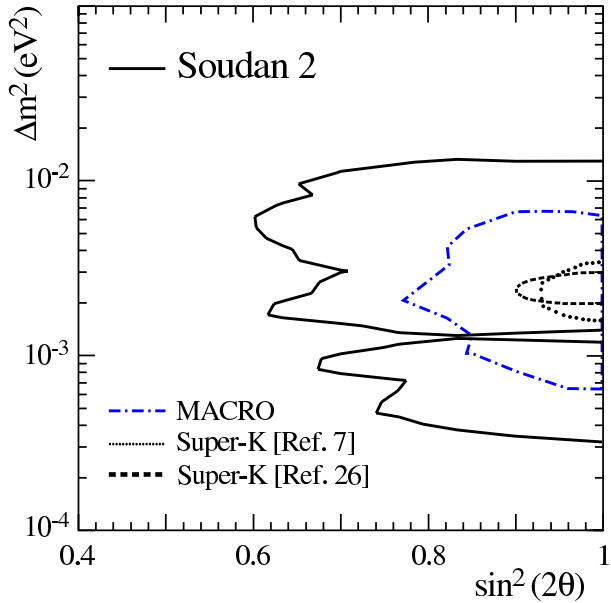,width=8.0cm}} 
\caption{ The Soudan 2 90\% confidence allowed region in \sqdm
(solid line) compared with the allowed regions of  MACRO
(dot-and-dashed line) \cite{MACROb}, and of the Super-K
zenith angle \cite{SK-USUT} (dotted line) and
$L/E$ \cite{SuperKallowed} (dashed line) analyses.}
\label{fig:comp}
\end{figure}
   
Samples of upward stopping muons produced by neutrino interactions
in the rock below the Soudan-2 detector and 
partially-contained events with downward-going
muons produced in the detector have been separately isolated.
These two new data sets provide additional support and constraints
for the hypothesis of atmospheric-neutrino oscillations.  
The flux of upward stopping neutrino-induced muon events is
observed to be suppressed by a factor of approximately two, 
while downward-going muon
events are not suppressed.  An oscillation analysis using the method
described in Ref. \cite{msanchez} and adding this new data gives a
more restrictive 90\%-confidence-level allowed region of \delm and \sqsin.
The probability of the no-oscillation hypothesis is reduced by more than
a factor of 10, to $3.2\times10^{-5}$.

The data have been analyzed using three models of the atmospheric flux
at the northern geomagnetic latitude of this experiment. The models include
two recent three-dimensional flux calculations and an older one-dimensional
calculation. The oscillation parameters are found to be essentially
independent of the flux calculation. 
The normalization factor for the experiment, 0.91$\pm$0.07, is the 
measured event rate divided by the calculated event rate 
where the latter is the convolution of neutrino fluxes of the
Bartol 3D flux calculation with neutrino cross sections 
encoded in NEUGEN3. The denominator for this
ratio contains elements which are specific to the Soudan-2 detector
analysis, and site.  Consequently, the normalization factor
cannot be compared in a straightforward way to other
experiments at other geomagnetic latitudes with different detector
media and using different neutrino interaction generators.

Comparison of this experiment's revised 90\% CL allowed region 
with the most recent Super-K \cite{SK-USUT, SuperKallowed} and
MACRO \cite{MACROb} allowed regions is shown in Fig.~\ref{fig:comp}. 
This result is in good agreement with both experiments.

\section*{Acknowledgments} 

   This work was supported by the U.S. Department of Energy, the U.K.
Particle Physics and Astronomy Research Council, and the State and
University of Minnesota.  We gratefully acknowledge the Minnesota Department
of Natural Resources for allowing us to use the facilities of the Soudan
Underground Mine State Park.  We warmly thank the Soudan 2 mine crew for
their dedicated work throughout the duration of the experiment.


\begin{thebibliography}{99} 

\bibitem{msanchez} Soudan 2 Collaboration, 
 M. Sanchez {\it et al.}, Phys. Rev. D {\bf 68},
 113004 (2003).

\bibitem{MACRO-USID} MACRO Collaboration, 
 M. Ambrosio {\it et al.},
 Eur. Phys. J. C {\bf 36}, 323 (2004).

\bibitem{Kam-UT} Kamiokande Collaboration, Y. Oyama {\it et al.}, 
Phys. Rev. D {\bf 39}, 1481 (1989);
 S. Hatakeyama {\it et al.}, Phys. Rev. Lett. {\bf 81}, 2016 (1998).

\bibitem{IMB-USUT} IMB Collaboration, 
 R. Becker-Szendy {\it et al.}, Phys. Rev. Lett. {\bf 69}, 1010 (1992);
 R. Clark {\it et al.}, Phys. Rev. Lett. {\bf 79}, 345 (1997).

\bibitem{SK-UT} Super-Kamiokande Collaboration, Y. Fukuda {\it et al.},
 Phys. Rev. Lett. {\bf 82}, 2644 (1999); Phys. Rev. Lett. {\bf 85}, 3999 (2000).

\bibitem{Frejus-USID} Frejus Collaboration, K. Daum {\it et al.},
 Z. Phys. C {\bf 66}, 417 (1995).

\bibitem{SK-USUT} Super-Kamiokande Collaboration, Y. Ashie {\it et al.},
 eprint hep-ex/0501064, submitted to Phys. Rev. D.

\bibitem{S2:NIM_A376_A381} Soudan 2 Collaboration, W.W.M. Allison 
 {\it et al.}, Nucl. Instr. Meth. A {\bf 376}, 36 (1996); 
 Nucl. Instr. Meth. A {\bf 381}, 385 (1996). 

\bibitem{S2:NIM_A276} Soudan 2 Collaboration, W.P. Oliver 
 {\it et al.}, Nucl. Instr. Meth. A {\bf 276}, 371 (1989). 

\bibitem{PDK-803} T. Kafka, H.R. Gallagher, W.A. Mann, J.K. Nelson, 
 Soudan-2 note PDK-803, June 2003.

\bibitem{bartol89} G. Barr, T.K. Gaisser and T. Stanev,
Phys. Rev. D {\bf 39}, 3532 (1989).

\bibitem{bartol96} V. Agrawal, T.K. Gaisser, P. Lipari and T. Stanev,
 Phys. Rev. D {\bf 53}, 1314 (1996).

\bibitem{Bartol04} G.D. Barr, T.K. Gaisser, 
P. Lipari, S. Robbins, and T. Stanev,
Phys. Rev. D {\bf 70}, 023006 (2004).

\bibitem{Battistoni_3D} G. Battistoni {\em et al.}, Astropart. Phys. 
 {\bf 12}, 315 (2000); G. Battistoni, A. Ferrari, T. Montaruli, and P.R. 
 Sala, eprint astro-ph/0207035, July 2002.
 Flux tables for the Soudan site are available 
 at http://www.mi.infn.it/battist/neutrino.html.

\bibitem{neugen} G. Barr, D.Phil thesis, University of Oxford, 1987;
H.M. Gallagher and M.C. Goodman, MINOS note NuMI-112, Nov. 1995;
H. Gallagher, Nucl. Phys. B (Proc. Suppl.) {\bf 112}, 188 (2002).

\bibitem{demuth} Soudan~2 Collaboration, D. Demuth {\it et al.},
 Astropart. Phys. {\bf 20}, 533 (2004);
 D. M. Demuth, PhD thesis, University of Minnesota, 1999.

\bibitem{Feldman_Cousins} G.J. Feldman and R.D. Cousins, 
 Phys. Rev. D {\bf 57}, 3873 (1998).

\bibitem{Ruddick} H. Gallagher and K. Ruddick, 
 Soudan-2 note PDK-784, Jan. 2002.
 
\bibitem{G-Garcia_RMP03} M.C. Gonzalez-Garcia and Y. Nir, 
Rev. Mod. Phys. {\bf 75}, 345 (2003).
 
\bibitem{PDK-811} W.A. Mann, K. Galdamez, T. Kafka, and A. Sousa,  
 Soudan-2 note PDK-811, Sept. 2004.

\bibitem{Peres-Smirnov} O.L.G. Peres and A.Yu. Smirnov, 
 Phys. Lett. B {\bf 456}, 204 (1999);
 Nucl. Phys. B (Proc. Suppl.) {\bf 110}, 355 (2002).

\bibitem{PDK-807} W.A. Mann and M. Roberto, Soudan-2 note
PDK-807, July 2003.

\bibitem{Kamland} KamLAND Collaboration, T. Araki {\it et al.},
 Phys. Rev. Lett. {\bf 94}, 081801 (2005).
 
\bibitem{Chung} Soudan 2 Collaboration, 
 J. Chung {\it et al.}, Phys. Rev. D {\bf 66},
 032004 (2002).
 
\bibitem{PDK-791} R.T. Thompson, W.A. Mann, and M.A. Said, Soudan-2 note
PDK-791, August 2002.
 
\bibitem{SuperKallowed} Super-Kamiokande Collaboration, Y. Ashie {\it et al.},
 Phys. Rev. Lett. {\bf 93}, 101801 (2004).
 
\bibitem{MACROb} MACRO Collaboration, M. Ambrosio {\it et al.},
 Phys. Lett. B {\bf 566}, 35 (2003).


\end{thebibliography}
\end{document}